\documentclass[conference]{IEEEtran}

\usepackage{algorithm}
\usepackage{algpseudocode}
\usepackage{amsmath}
\usepackage{amssymb}
\usepackage{color}
\usepackage{graphicx}

\usepackage{multirow}
\usepackage{booktabs} 
\usepackage{array}
\usepackage{caption}
\usepackage{xcolor}

\usepackage[normalem]{ulem}

\ifCLASSINFOpdf

\else

\fi

\begin{document}

\title{
Redefining DDoS Attack Detection Using A Dual-Space Prototypical Network-Based Approach
}

\author{\IEEEauthorblockN{Fernando Martinez}
\IEEEauthorblockA{Computer \& Information Science Dept.\\
Fordham University\\
NY, USA 10023\\
fmartinezlopez@fordham.edu}
\and
\IEEEauthorblockN{Mariyam Mapkar}
\IEEEauthorblockA{Computer \& Information Science Dept.\\
Fordham University\\
NY, USA 10023\\
mmapkar@fordham.edu}
\and
\IEEEauthorblockN{Ali Alfatemi}
\IEEEauthorblockA{Computer \& Information Science Dept.\\
Fordham University\\
NY, USA 10023\\
aalfatemi@fordham.edu}
\and
\IEEEauthorblockN{Mohamed Rahouti}
\IEEEauthorblockA{Computer \& Information Science Dept.\\
Fordham University\\
NY, USA 10023\\
mrahouti@fordham.edu}
\and
\IEEEauthorblockN{Yufeng Xin}
\IEEEauthorblockA{RENCI\\
UNC at Chapel Hill\\
NC, USA 27517\\
yxin@renci.org}
\and
\IEEEauthorblockN{Kaiqi Xiong}
\IEEEauthorblockA{Cyber Florida\\
University of South Florida\\
FL, USA 33617\\
xiongk@usf.edu}
\and
\IEEEauthorblockN{Nasir Ghani}
\IEEEauthorblockA{Electrical Engineering Dept.\\
University of South Florida\\
FL, USA 33617\\
nghani@usf.edu}
}

\maketitle

\begin{abstract}
Distributed Denial of Service (DDoS) attacks pose an increasingly substantial cybersecurity threat to organizations across the globe. In this paper, we introduce a new deep learning-based technique for detecting DDoS attacks, a paramount cybersecurity challenge with evolving complexity and scale. Specifically, we propose a new dual-space prototypical network that leverages a unique dual-space loss function to enhance detection accuracy for various attack patterns through geometric and angular similarity measures. This approach capitalizes on the strengths of representation learning within the latent space (a lower-dimensional representation of data that captures complex patterns for machine learning analysis), improving the model's adaptability and sensitivity towards varying DDoS attack vectors. Our comprehensive evaluation spans multiple training environments, including offline training, simulated online training, and prototypical network scenarios, to validate the model's robustness under diverse data abundance and scarcity conditions. The Multilayer Perceptron (MLP) with Attention, trained with our dual-space prototypical design over a reduced training set, achieves an average accuracy of 94.85\% and an F1-Score of 94.71\% across our tests, showcasing its effectiveness in dynamic and constrained real-world scenarios.

\end{abstract}


\IEEEpeerreviewmaketitle

\section{Introduction} \label{sec:introduction}

Distributed Denial of Service (DDoS) attacks represent a significant and growing cybersecurity challenge, disrupting operations within organizations worldwide. Traditional detection systems struggle to keep pace with evolving attack methods, underscoring the need for innovative solutions that adapt to the dynamic nature of cybersecurity threats \cite{guo2023review, wei2022abl}. This research introduces a groundbreaking approach: a dual-space prototypical network that leverages geometric and angular measures in a dual-space framework, effectively enhancing the detection and adaptability to both known and novel DDoS patterns \cite{debicha2023tad}.

The evolution of DDoS attacks in terms of scale and complexity has reached a point where they pose a credible threat to large corporations and governmental bodies alike \cite{mirkovic2004taxonomy}. The severity of the issue is highlighted by record-breaking incidents, including the Mirai botnet's 620 Gbps assault on Dyn in 2016, leading to widespread disruption of services for major platforms such as Twitter, Netflix, and Spotify \cite{kolias2017ddos}. Furthermore, in 2018, GitHub became the target of an unprecedented 1.7 Tbps DDoS attack, marking it the most significant attack of its kind to date \cite{newman20181}. These examples underscore the urgent need for robust mechanisms to promptly identify and mitigate DDoS attacks, safeguarding organizational assets and maintaining the integrity of digital infrastructures \cite{rahouti2022sdn}.

In the domain of DDoS attack detection, the state-of-the-art has evolved significantly to address these threats' increasing complexity and frequency. Recent advancements have focused on leveraging machine learning (ML) and deep learning techniques to improve detection accuracy and response times \cite{guo2023review}. These approaches utilize extensive datasets to train models capable of identifying patterns indicative of DDoS activities, distinguishing them from legitimate network traffic. Deep learning, with its ability to learn complex patterns from data, has shown particular promise in this area. Models such as Convolutional Neural Networks (CNNs), Recurrent Neural Networks (RNNs), and variations like Long Short-Term Memory (LSTM) networks have been applied to parse and analyze network traffic, offering improved precision in identifying DDoS attacks over traditional signature-based and anomaly-based detection systems.

Previous methods, such as few-shot learning, were limited by their reliance on static models, conventional data analysis techniques, and a single-space representation. This approach restricted the adaptability and accuracy of detection systems, as it did not fully leverage the potential of spatial analysis or consider the complex, evolving nature of cybersecurity threats. The absence of dynamic adjustment mechanisms further constrained these systems' effectiveness against new or evolving DDoS attacks, making them less resilient in the face of sophisticated cybersecurity challenges.

Further, researchers have explored adaptive learning models and transfer learning techniques to enhance the detection framework's responsiveness to new and evolving DDoS threats \cite{wei2022abl, dhillon2021building}. Adaptive models are designed to continuously learn and adjust to new data, enabling them to identify abnormal patterns that could indicate an attack \cite{guo2023review}. Transfer learning allows these models to apply knowledge gained from detecting known attacks to identify new, related threats, thereby reducing the need for extensive retraining with new data \cite{debicha2023tad}. This shift towards models that can dynamically adapt to changing attack patterns represents a critical advancement in the field, setting the foundation for more resilient cybersecurity defenses against the ever-evolving landscape of DDoS threats.

This research represents a step forward by introducing the dual-space loss function within prototypical networks to enhance DDoS attack detection. Our framework not only encapsulates deep learning techniques but also tailors them specifically for the task of identifying cyber threats. By incorporating multifaceted data, our model gains a detailed understanding of network behaviors, allowing it to pick up on the subtle signs that might indicate an attack. The adaptive learning angle of our algorithm is tailored to dynamically adjust to changing attack patterns, harnessing previously garnered insights to refine its detection accuracy. This approach eliminates the extensive retraining typically required, ensuring our model remains adept and responsive in the face of new DDoS threats. The deployment of our algorithm in real-world cybersecurity operations offers a significant leap forward in safeguarding digital infrastructures. By providing a robust mechanism capable of rapidly adapting to and detecting both existing and emerging DDoS threats, our approach directly addresses the critical need for dynamic defense strategies. This innovation not only enhances the security posture of organizations but also significantly reduces the time and resources required for incident response, showcasing our research's practical implications and tangible benefits. 

Our empirical evaluation underscores the consistent superiority of the dual-space prototypical network, particularly when navigating the complexities of limited labeled data availability compared to conventional deep learning approaches or even regular prototypical networks and showcasing an average F-1 score and accuracy of 94.71\% and 94.85\%, respectively, across all our experiments, ensuring empirical robustness.

The key contributions of this paper are summarized as follows.
\begin{itemize}
    \item Introduction of a dual-space prototypical network, a deep learning model designed for identifying DDoS attacks. It employs a unique dual-space loss function to enhance detection accuracy for both known and new attack patterns by leveraging geometric and angular similarity measures.
    \item Application of the dual-space loss function, exploiting the power of representation learning in the latent space, to improve the model's adaptability and sensitivity to known and unseen DDoS attack vectors. Adaptability refers to the model's internal mechanism for handling variability in attack patterns.
    \item Comprehensive evaluation across various training environments, including offline training, simulated online training, and prototypical network scenarios, to showcase the model's robust performance under data abundance and scarcity conditions. The key evaluation results indicate that the model, particularly the dual-space prototypical network with attention trained on a reduced training set, achieved high accuracy (94.85\%) and an F1-Score (94.71\%) on average, demonstrating its robustness in dynamic real-world scenarios.
    \item Demonstration of the framework's adaptability (i.e., the model's external validation to validate its versatility and effectiveness in real-world conditions) and learning efficiency through rigorous testing against diverse learning paradigms and data availability conditions, establishing a new benchmark for adaptive defense strategies against DDoS threats.
\end{itemize}

The rest of this paper is structured as follows. Section \ref{sec:related} reviews the state-of-the-art related to this work. Next, Section \ref{sec:problem} discusses the research problem tackled in this work, and Section \ref{sec:methodology} details our proposed methodology, framework design, and data processing. The key results are presented next in Section \ref{sec:evaluation} to demonstrate the efficiency of the proposed methodology. Lastly, Section \ref{sec:conclusion} concludes this paper and highlights future plans.

\section{Related Work} \label{sec:related}

The rise in DDoS attacks has prompted researchers and service providers to dedicate more effort towards exploring and countering these incidents \cite{David2021Discriminating}. Conventional detection methods at the network layer, like analyzing packet headers or traffic flow patterns, might not be adequate for effectively identifying and mitigating contemporary DoS/DDoS attacks. This challenge has led to a significant increase in research and the development of new detection and mitigation techniques \cite{Chin2018Kernel, Zheng2018Realtime, Rahouti2021SynGuard, Liang2019Empirical, owusu2023enhancing, alfatemi2024advancing}.

Recent advancements in AI have propelled the use of deep learning in intrusion detection, significantly enhancing detection efficiency and effectiveness. Pioneering works like Zhang et al. \cite{zhang2019network} have utilized raw data to minimize information loss, integrating enhanced LeNet-5 and LSTM for superior results. Similarly, Zhong et al. \cite{zhong2020helad} combined LSTM with AutoEncoder to compute abnormal scores for network traffic flagging, demonstrating notable improvements over traditional methods like SVM, IF, and GMM. Further contributions by Li et al. \cite{li2022mfvt} and Wei et al. \cite{wei2022abl} have incorporated PCA for raw traffic feature extraction and attention-based LSTM models for improving detection accuracy, respectively. These developments underscore a significant shift towards leveraging complex models to address the intricacies of network intrusion detection, highlighting the critical role of deep learning in identifying and mitigating malicious activities within network environments.

Notable studies like Sun et al.~\cite{sun2023few} introduced a few-shot network intrusion detection approach utilizing a Prototypical Capsule Network with an attention mechanism, enhancing the ability to capture spatial feature hierarchies. Similarly, Miao et al.~\cite{miao2023spn} developed SPN, a Siamese Prototypical Network-based method for traffic classification that incorporates out-of-distribution detection, aiming to enhance generalization from limited examples. Hosseini also \cite{hosseini2023intrusion} studied intrusion detection using few-shot learning, addressing the adaptability challenges in resource-constrained networks. Tian et al.~\cite{tian2022few} proposed an enhanced parallelized triplet network for network intrusion detection, focusing on improving feature discriminative power.

Iliyasu et al.~\cite{iliyasu2022few} presented a system using discriminative representation learning with a supervised autoencoder for few-shot network intrusion detection, aiming at refining feature extraction for sophisticated attack detection. Yang et al.~\cite{yang2022fs} developed FS-IDS, a framework for intrusion detection, emphasizing adaptability and scalability to maintain detection capabilities with scarce training data.

Furthermore, Yu et al. \cite{yu2020intrusion} introduced an intrusion detection system (IDS) model leveraging the few-shot learning concept, employing CNN and DNN to highlight essential features. A separate study by Kwon et al. \cite{kwon2018empirical} explored the use of Variational Autoencoders (VAEs), Fully Connected Networks (FCNs), and Sequence-to-Sequence (Seq2Seq) models for identifying network anomalies. Wang et al. \cite{wang2018network} developed a prediction model for network flow, combining deep learning with ensemble learning strategies. Various deep learning architectures, such as Long Short-Term Memory (LSTM) and Convolutional Neural Networks (CNN), have been employed to boost the effectiveness of IDS systems \cite{jiang2020network, liu2021fast}. Li et al.~\cite{li2019method} proposed a method that combines LSTM to analyze temporal characteristics and CNN to assess spatial features for detecting HTTP malicious traffic in mobile networks.

This study introduces a new approach to detecting DDoS attacks using representation learning in a dual-space framework. Unlike previous methods that rely mainly on static models, conventional data analysis techniques, and single-space representation for few-shot learning scenarios, our approach incorporates geometric and angular information to improve detection accuracy. We also use an attention mechanism to adapt to new threats dynamically. Our work demonstrates the importance of spatial analysis and dual-dimensional insights in developing advanced and resilient DDoS detection systems. It establishes a new precedent for deep learning applications in cybersecurity.

\section{Research Problem} \label{sec:problem}

The research problem addressed in this paper is the increasing prevalence and sophistication of DDoS attacks, which pose significant challenges to cybersecurity. Traditional defense mechanisms, which rely on historical networking traffic, are sometimes inefficient due to their lack of flexibility and inability to adapt to attackers' novel and evolving tactics. This limitation is particularly critical in pattern-changing attacks, where attackers continuously adjust their strategies, rendering static detection methods ineffective. The proposed work seeks to overcome these shortcomings by developing a deep learning framework that leverages a novel dual-space prototypical network. 

Further, the essence of our research lies in addressing the dual challenges of accuracy and adaptability in DDoS attack detection. By leveraging a dataset encompassing network traffic, server logs, system events, and user activities, we aim to define a detection system that is both comprehensive and dynamic. We rigorously evaluated our method against diverse learning paradigms, including offline training, a simulated online training environment, and a standard prototypical network approach. This evaluation spanned scenarios characterized by both an abundance of labeled data and conditions of limited data availability, specifically using sets of only 100 samples for training. Such exhaustive testing was designed to scrutinize our method's adaptability and performance across a spectrum of training conditions, thus demonstrating its robustness and capacity to adjust to data-rich and data-constrained environments.

This work seeks to bridge the critical gap in existing defense methodologies by developing a deep learning framework that introduces a dual-space prototypical network. This network provides an efficient approach to DDoS attack detection, integrating within Multilayer Perceptron (MLP) architectures to achieve adaptive and precise detection by leveraging episodic learning from a support and query set, enhancing the model's ability to classify new, unseen samples. Our approach outperforms the limitations of record-based systems by harnessing the power of adaptive representation learning. By integrating MLP models with our dual-space loss function, our framework distinguishes between recognizing established attack vectors and adapting to new, unidentified patterns. This dual-space loss function, combining normalized Euclidean and cosine distances, enables our model to evaluate the similarity between data points more holistically—considering not just the magnitude of feature differences but also their directional alignment in the latent space.

Last, regarding the choice of binary classification over multi-class classification, this work primarily focuses on distinguishing between normal (benign) network traffic and malicious DDoS traffic. This binary distinction is critical in intrusion detection, where the immediate goal is to effectively detect and mitigate DDoS attacks. While multi-class classification can offer insights into different types of DDoS attacks, the primary objective here is the rapid and accurate identification of DDoS activities to safeguard organizational assets and maintain the integrity of digital infrastructures.

\section{Methodology} \label{sec:methodology}

This section presents our methodology, which includes a description of the dataset used, preprocessing transformations, and the foundations of our proposed methodology. Our approach involves using two neural network architectures - a foundational MLP and an enhanced version with an attention mechanism. The dual-space prototypical network is a unique component within our two neural network architectures. This network disregards the conventional output layer and employs the last hidden layer as the functional output. Our network can intricately map and recognize DDoS attacks due to the optimization of our dual-space loss function, enabling informed strategic reframing of the neural architecture and enhanced performance (especially in sparsely labeled data).

Prototypical networks are chosen for DDoS detection in this work due to their advantages in few-shot learning, allowing effective classification with limited training examples \cite{xu2020method}. This feature is crucial for intrusion detection, where data on new threats may be scarce. Unlike traditional models that need large datasets for high accuracy, prototypical networks quickly adapt to new patterns, making them ideal for the rapidly evolving landscape of DDoS threats. They simplify the learning process by focusing on a metric space for classification, reducing complexity and enhancing interpretability. Their suitability for detecting new attack vectors with minimal data underpins their selection for addressing the dynamic challenge of DDoS attack detection.

\subsection{Framework and Design Foundations}


The foundation of our approach lies in the design of our deep learning framework and how we update its weights. Central to this design is the preprocessing of data, a step that lays the groundwork for subsequent layers of analysis. We utilize RobustScaler to handle the diverse nature of cybersecurity data, diligently scaling features by centering them around the median and normalizing based on the Interquartile Range (IQR). As a result, it imparts resilience against outliers that are commonly found in cyber attack datasets. The proposed framework for few-shot DDoS detection is expressed by Algorithm \ref{alg:alg1} and detailed next.

\begin{equation}
\text{\textit{Scaled feature}} = \frac{\text{\textit{feature}} - \text{\textit{median}}}{IQR}
\end{equation}

One of our proposed architecture's critical components is the attention-based mechanism. This part of the model enables it to focus sharply on the most important features among a large number of attributes. The attention mechanism is not just an additional feature but a revolutionary approach to handling tabular data within neural networks. The attention mechanism steps are formulated in Table \ref{table:attention_mechanism}.

\begin{table}[ht]
\centering
\begin{tabular}{ll}
\hline
\textbf{Step}                & \textbf{Formula}                                   \\ \hline
Linear Projection           & $projection = Wx$                                  \\
Tanh Activation             & $tanh\ projection = \tanh(projection)$             \\
Attention Weights           & $attention = \frac{e^{tanh\_projection}}{\sum e^{tanh\_projection} + \epsilon}$ \\
Weighted Input              & $weighted\ input = x \cdot attention$              \\ \hline
\end{tabular}
\caption{Summary of the attention mechanism steps.}
\label{table:attention_mechanism}
\end{table}

In our efforts to create robust DDoS attack detectors, we developed an improved Prototypical Network that integrates a novel dual-space loss, leveraging both Euclidean and cosine distance measures. Our approach offers distinct advantages, enhancing our model's ability to recognize and adapt to new, unseen attacks, even with limited training samples, while remaining efficient. Initially introduced by Snell et al.~\cite{snell2017prototypical}, Prototypical Networks excel in few-shot learning environments by learning to represent each class with a prototype- a method particularly effective when data is scarce. The training uses episodic scenarios, each mimicking a few-shot learning challenge, thereby cultivating a resilient and adaptable model. The cornerstone of this approach is the prototypical loss \(\mathcal{L}_{\text{prot}}\), which traditionally measures the distance between query embeddings and their respective class prototypes.

\begin{equation*}
\mathcal{L}_{\text{prot}} = - \sum_{(x_j,y_j) \in Q} \log \frac{\exp(-d(q_j, p_{y_j}))}{\sum_{c \in C} \exp(-d(q_j, p_c))}
\end{equation*}

\subsection{Dual-space Loss}
We designed the dual-space loss (denoted as $\mathcal{L}_{\text{D-space}}$) to capture both the geometric proximity between embeddings and prototypes and their directional alignment, offering a more sophisticated understanding of class similarities. This intricate measure is crucial in few-shot learning when obtaining labeled examples of attacks is often impractical due to the evolving nature of threats. Our experiments underscore the method's effectiveness, revealing significant performance enhancements across various attack scenarios, even when operating with small training sets.

\begin{equation}
\mathcal{L}_{\text{D-space}} = -\sum_{i=1}^{N} \log \frac{\exp(-D_{\text{E+C}, i, y_i})}{\sum_{j=1}^{C} \exp(-D_{\text{E+C}, i, j})}
\label{eq:dual_space_loss}
\end{equation}
\begin{equation}
D_{\text{E+C}, i, j} = \alpha \cdot D_{\text{E}, i, j} + (1-\alpha) \cdot D_{\text{C}, i, j}
\label{eq:combined_distance}
\end{equation}
\begin{equation}
D_{\text{E}, i, j} = \frac{\|\mathbf{q}_i - \mathbf{p}_j\|_2}{\sum_{k=1}^{C} \|\mathbf{q}_i - \mathbf{p}_k\|_2}
\label{eq:normalized_euclidean_distance}
\end{equation}
\begin{equation}
D_{\text{C}, i, j} = 1 - \frac{\mathbf{q}_i \cdot \mathbf{p}_j}{\|\mathbf{q}_i\|_2 \|\mathbf{p}_j\|_2}
\label{eq:cosine_distance}
\end{equation}

Where:
\begin{itemize}
    \item $\alpha$: Weighting factor that balances the influence of the normalized distances in our dual-space loss.
    \item $N$ The number of query samples.
    \item $C$: The number of classes represented in the support set $S$.
    \item $q_i$: The embedding of the $i$-th query sample.
    \item $p_j$: The prototype of the $j$-th class.
    \item $\|\cdot\|_2$: The L2 norm (euclidean norm) of a vector used here to calculate Euclidean distances and to normalize vector for cosine similarity.
    \item $D_{\text{E}, i, j}$: Normalized Euclidean distance between the $i$-th query sample and the $j$-th class prototype.
    \item $D_{\text{C}, i, j}$: Cosine distance between the $i$-th query sample and the $j$-th class prototype, calculated as 1 \textit{minus} the cosine similarity.
    \item $D_{\text{E+C}, i, j}$: Combined weighted distance of both normalized Euclidean and cosine distance.
\end{itemize}

The presented loss evaluation-specific metric, $\mathcal{L}_{\text{D-space}}$, is leveraged to exploit the complementary strengths of geometric and angular similarity measures. The normalized Euclidean distance delivers a scale-invariant metric that quantifies absolute differences in attack signatures, addressing the variability in attack magnitudes without being overshadowed by outlier values. This aspect is particularly crucial when attack payloads vary dramatically but still belong to the same attack pattern (Zargar et al. \cite{pittir}). Further, cosine distance captures the angular proximity between embeddings, emphasizing the pattern similarity or dissimilarity irrespective of the vector magnitudes. This way, our function catches subtle behavioral patterns characteristic of sophisticated DDoS attacks, a technique echoed in works by Papernot et al. \cite{pepernot}, who stressed the importance of understanding directional tendencies in adversarial attack vectors.

Experimental results demonstrate significant performance improvements over traditional approaches, further affirming the effectiveness of our methodology, as we will see in section \ref{sec:evaluation}. Leveraging a dual-space loss allows a more sophisticated understanding of class similarities, enhancing the model's generalization ability from a few DDoS training instances.

\begin{table}[h]
\centering
\begin{tabular}{|p{2.5cm}|p{5.2cm}|} \hline
Feature & Info  \\ 
 \hline
Decision tuple & ID, src/dest IP, src/dst port, protocol \\ \hline
Time &  Timestamp, duration \\ \hline
Fwd pkts & Total, len (total, max, min, std, min) \\ \hline
Bwd pkts & Total, len (total, max, min, std, min)  \\ \hline
IAT & Mean, std, max, min \\ \hline
Fwd IAT & Total, mean, std, max, min \\ \hline
Bwd IAT & Total, mean, std, max, min \\ \hline
Fwd flags & Push, URG \\ \hline
Flags & Bwd (Push, URG), Count \\ \hline
Pkts len/size & Pkts (min, std, max, mean, var), size (avg) \\ \hline
Pkt loss & Down/up ratio \\ \hline
Flags count & FIN/SYN/RST/PSH/ACK/URG/ CWE/ECE   \\ \hline
Fwd pkt header & Len, avg (Seg size, bytes/bulk, bulk rate) \\ \hline
Bwd pkt header & Len, avg (Seg size, bytes/bulk, bulk rate)     \\ \hline
Subflow & Fwd/bwd (pkts, Bytes) \\ \hline
Init win Bytes & Fwd, Bwd    \\ \hline
Active/idle & Mean, max, std, min  \\ \hline
Other labels & Inbound, Similar HTTP \\ \hline
\end{tabular}
\caption{Flow features and statistics in the CICIDS dataset.}
\label{tab:data-features}
\end{table}

\subsection{Dataset and Data Preprocessing} \label{subsec:data_collection}

In this paper, we employed the detailed CIC dataset referenced by Sharafaldin et al. \cite{sharafaldin2018toward}, meticulously compiled using the \textit{CICFlowMeter} tool \cite{CICFlowMeter}. These features are then preprocessed to normalize the data formats and remove irrelevant or noisy data. This preprocessing includes normalization, tokenization, and data cleansing to ensure the datasets are ready for effective feature extraction. This dataset, covering several days, is rich in network flow data, each entry being detailed with 83 unique attributes and comprising a total of 225,745 entries.

Noteworthy attributes listed in Table \ref{tab:data-features} include \texttt{flag\_rst}, highlighting the Reset flag in TCP headers; \texttt{pk\_len\_std}, showing the standard deviation of packet lengths; \texttt{fwd\_subflow\_bytes\_mean}, indicating the average byte size of forward subflows; \texttt{flow\_duration}, capturing the total duration of network flows; \texttt{bwd\_pkt\_len\_mean}, representing the average size of backward packets; and \texttt{bwd\_pkt\_len\_tot}, which accounts for the total length of packets moving in the opposite direction. This dataset provides an in-depth view of network behaviors and patterns, proving invaluable for our analysis.

\subsubsection{Feature extraction and selection} \label{subsec:feature_extraction}

We employ sophisticated algorithms to extract meaningful variables from each data source. For network traffic, we focus on extracting features like packet size, flow duration, and protocol type. Server logs are analyzed for patterns indicating unauthorized access or system misuse. System event logs provide insights into unusual system behavior, and user activity logs are scrutinized for patterns deviating from normal behavior. 

To refine our feature set further and ensure our model focuses on the most predictive indicators of attacks, we employ a feature selection process leveraging an initial Random Forest algorithm. This model is trained on the entirety of the dimensions, utilizing its inherent bagging technique to reduce overfitting while capturing the underlying structure of the data. Post-training, we harness the Random Forest's feature importance scores to filter the feature set down to those with the highest relevance to attack detection. This process boosts the model's performance by focusing on the most informative features and enhances model efficiency by eliminating redundant or less significant attributes. 

\subsection{Model Design} \label{subsec:model_development}

\subsubsection{MLP and attention mechanisms}

The subsequent section expounds on developing two distinctive neural network architectures: a foundational MLP and an augmented variant incorporating an attention mechanism. Notably, these architectures are not standalone entities but are integral components of our broader strategy to enhance the detection of DDoS attacks.

\paragraph{MLP foundation} At its core, our approach utilizes the MLP model's robust feature-extraction capabilities. Structured through sequential layers, the MLP efficiently distills complex patterns from input data, transitioning from simple pattern recognition in initial layers to intricate feature synthesis in deeper tiers. This hierarchical processing is pivotal for differentiating between benign and malicious network activities. To bolster the MLP's efficacy in DDoS detection, we've integrated dropout layers to mitigate overfitting, employed batch normalization for faster convergence, and adopted ReLU activation functions to enhance non-linear processing capabilities.

\paragraph{MLP with attention mechanisms} Building on the MLP's solid foundation, we have infused our architecture with the attention mechanism we defined in Table \ref{table:attention_mechanism}, drawing inspiration from seminal works by Kim et al.\cite{kim2017structured}. These mechanisms endow the model with a dynamic focus capability, enabling it to prioritize analyzing input segments most indicative of DDoS attack instances. By computing and applying relevance scores across the input data, the model highlights features crucial for identifying obvious and subtle DoS threats, including infrequent attacks. This adaptive focus enables our model to assign importance scores to different inputs autonomously, or guided attention, where external knowledge or signals direct the model's focus.

\paragraph{Dual-space prototypical network integration} The dual-space prototypical network (\textit{D-space}) represents a leap in our capacity to model and understand the complex data representations associated with DDoS attacks. While demonstrated through our MLP architectures, this framework is inherently model-agnostic, underscoring its versatility and broad applicability. It can seamlessly integrate with any neural network architecture capable of yielding an embedding output, as represented by $f_\phi(x_v)$ in Algorithm \ref{alg:alg1}. Similar to regular prototypical networks, our approach involves learning from episodic batches. Each \textit{episode} is constructed from a support set and a query set, simulating the challenge of learning from a few examples and then applying that knowledge to classify new, unseen samples.

\begin{algorithm}
\caption{Dual-space prototypical network for DDoS detection.}
\label{alg:alg1}
\begin{algorithmic}[1]
\State \textbf{Input}: Episodes $E = \{e_1, e_2,...,e_N\}$, where each episode $e$ contains:
\State \hspace{\algorithmicindent} Support set $S = \{(x_1, y_1), (x_2, y_2),..., (x_m, y_m)\}$
\State \hspace{\algorithmicindent} Query set $Q = \{(x_1, y_1), (x_2, y_2),..., (x_n, y_n)\}$
\State \textbf{Objective}: Learn embedding $f_\phi$ to minimize distance between embeddings of query samples and their corresponding class prototype.

\Procedure{Train Network}{}
    \For{each episode $e \in E$}
        \State \textit{// Calculate class prototypes}
        \For{each class $c \in C$}
            \State $p_c = \frac{1}{|S_c|} \sum_{(x_i, y_i) \in S_c} f_\phi(x_i)$
        \EndFor
        \State \textit{// Generate embeddings for query set}
        \For{each $(x_j, y_j) \in Q$}
            \State $q_j = f_\phi(x_j)$ 
        \EndFor

        \State \textit{// Dual-space prototype loss}
        \State $D_{\text{C}, j, c} = 1 - \frac{\mathbf{q}_j \cdot \mathbf{p}_c}{\|\mathbf{q}_j\|_2 \|\mathbf{p}_c\|_2}$
        \vspace{5pt}
        \State $D_{\text{E}, j, c} = \frac{\|\mathbf{q}_j - \mathbf{p}_c\|_2}{\sum_{k=1}^{C} \|\mathbf{q}_j - \mathbf{p}_k\|_2}$
        \vspace{5pt}
        \State $D_{\text{E+C}, j, c} = \alpha \cdot D_{\text{E}, j, c} + (1-\alpha) \cdot D_{\text{C}, j, c}$
        \vspace{5pt}
        \State $\mathcal{L}_{\text{D-space}} = -\sum_{(x_j,y_j) \in Q} \log \frac{\exp(-D_{\text{E+C}, j, y_j})}{\sum_{c=1}^{C} \exp(-D_{\text{E+C}, j, c})}$

        \vspace{5pt}     

        \State Update $\phi$ to minimize $\mathcal{L}_{\text{D-space}}$
    \EndFor
\EndProcedure
\State \textbf{return} Trained embedding function $f_\phi$

\Procedure{Estimate}{}
    \State \textbf{Input}: $X = \{x_1, x_2,...,x_V\}$
    \State \textit{// Generate embeddings}
    \State $q_v = f_\phi(x_v)$
    \State \textit{// Predict the closest prototype for each embedding}
    \vspace{5pt}  
    \State $d_{q_v, p_c} = \alpha \cdot \frac{\|\mathbf{q}_j - \mathbf{p}_c\|_2}{\sum_{k=1}^{C} \|\mathbf{q}_j - \mathbf{p}_k\|_2} + (1 - \alpha) \cdot (1 - \frac{\mathbf{q}_j \cdot \mathbf{p}_c}{\|\mathbf{q}_j\|_2 \|\mathbf{p}_c\|_2})$

    \vspace{5pt}  
    
    \State $\hat{y} = argmin(d_{q_v, p_c})$
\EndProcedure
\State \textbf{return} Estimated $\hat{y}$
\end{algorithmic}
\end{algorithm}

At the beginning of each episode, the algorithm calculates class prototypes by averaging the embeddings of samples within the support set for each class. In this step, the neural networks output a tangible representation of each class within the embedding space based on the available scarce examples. Then, for every sample in the query set, the model generates embeddings using the learned feature extraction capabilities of either the baseline MLP or the attention-augmented MLP. These embeddings are then compared against the class prototypes to calculate their dissimilarity and update the weights.


\section{Results} \label{sec:evaluation}


\subsection{Learning Techniques}
\subsubsection{Offline Learning}
In our study, offline training served as a foundational benchmark, allowing us to establish baseline performance metrics for our models under controlled, static conditions. This approach involved training the models on a pre-defined dataset to evaluate their learning efficacy and generalization capabilities before being subjected to the dynamic conditions of online learning or our prototypical-based approaches. 

\subsubsection{Online Learning}
Similar to the regular offline training scenario, we designed a simulated online training environment for benchmark purposes. Our design replicates real-world conditions where data is continuously received, necessitating continuous updates by the models. To replicate the constant data flow characteristic of real-time applications, we employed a generator function termed \textit{``data\_stream\_gen"} that simulates the chronological flow of real-world data streams by eschewing the conventional approach of random data shuffling. Instead, it maintains the natural order of data arrival.

Each iteration of our online learning process presents data to the model in small, sequentially ordered batches. This approach contrasts with the common practice of shuffling data and presenting it randomly. By adhering to a sequential order, we ensure that the learning process is continuously challenged by the evolving nature of the data, thereby providing a rigorous test of the model's capacity to adapt and update its parameters in real-time. Each batch triggers an update cycle within the model, comprising the forward pass, loss computation, backpropagation, and subsequent weight adjustment. 

\subsubsection{Prototypical and dual-space learning}
Within the ambit of our exploration, the Prototypical-based learning experiments are trained in an offline setting. In this way, our models aim to harness the full potential of the original prototypical approach and our dual-space contribution, learning robust representations of the data's embedding structure in a stable environment. 

\subsection{Consistent Experimental Setup} \label{expsetup}
To ensure the reliability and comparability of our findings, all experiments were conducted under a uniform experimental framework. This consistency was crucial across various dimensions:

\begin{itemize}
    \item \textbf{Neural network architectures:} We standardized the architecture for all our models to eliminate performance variances attributable to model complexity or capacity differences.
    \item \textbf{Optimizer:} The Adam optimizer was employed in all experiments, configured with a learning rate of $1 \times 10^{-3}$ and a weight decay of $0.01$. This choice was informed by the optimizer's proven efficiency in managing sparse gradients, as documented by Kingma and Ba \cite{kingma2017adam}, and its demonstrated effectiveness in our preliminary tests.
    \item \textbf{Feature preparation:} Before model training, we eliminated features that could lead to oversimplified or biased decision-making, such as source/destination IP addresses and ports. Employing the previously defined bagging-based feature selection technique, we curated a refined set of 28 crucial attributes.
    \item \textbf{Training duration and batch size:} For offline training scenarios, we adopted a batch size of 32 and a training duration of 10 epochs. This configuration was determined to be optimal for achieving a balance between training efficiency and model performance. In contrast, the online training setting required a continuous update mechanism without predefined epochs, reflecting the ongoing nature of real-time data processing. Thus, we set a maximum of 50 updates.
    \item \textbf{Experimental runs}: We conducted 30 runs for each configuration to ensure statistical significance. 
\end{itemize}

\subsection{Training Process Analysis On Reduced Training Set} \label{subsec:training-process-analysis}
\begin{figure}[htbp]
    \centering
    \includegraphics[scale=.50]{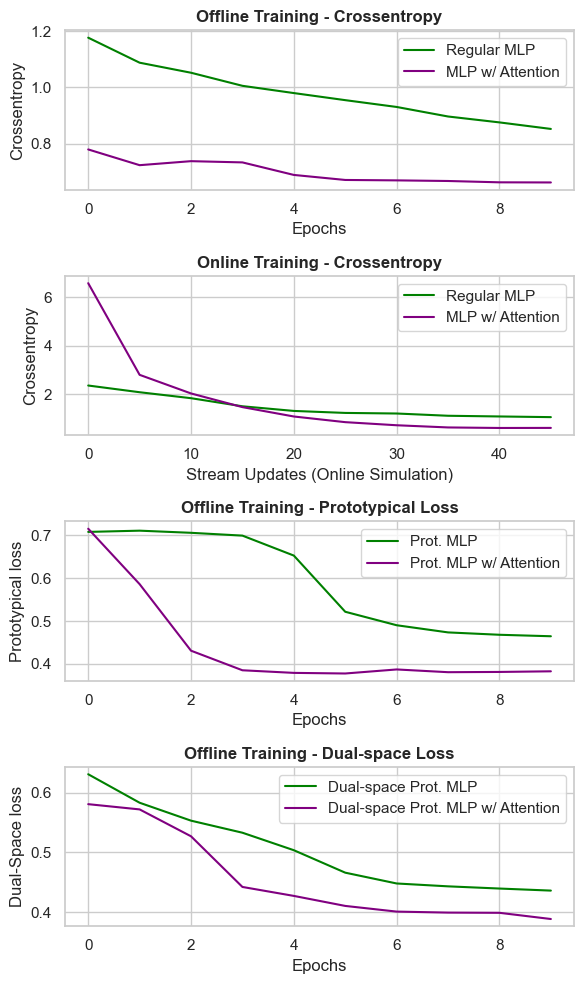}
    \caption{Validation losses during training on reduced dataset (Train $N$: 100).}
    \label{fig:lossfuncs}
\vspace{-6mm}
\end{figure}

In Figure \ref{fig:lossfuncs}, we observe the validation loss trends across different training paradigms on a reduced dataset (just 100 labeled samples). A striking observation is the comparative performance of the Offline Training with dual-space loss, which shows a consistent and robust decline in validation loss, indicative of effective learning and generalization from a limited number of training observations.

\begin{table*}[h]
\centering
\begin{tabular}{c|l|c|c|c|c}
\textbf{Model} & \multicolumn{1}{c|}{\textbf{\begin{tabular}[c]{@{}c@{}}Learning Technique\end{tabular}}} & \textbf{Accuracy} & \textbf{F-1} & \textbf{Precision} & \textbf{Recall} \\ \hline
\multirow{4}{*}{\textbf{MLP}} & Offline & $98.71 \pm 9.62e^{-2}$ & $98.69 \pm 9.7e^{-2}$ & $99.21 \pm 1.85e^{-1}$ & $98.19 \pm 2.34e^{-1}$\\ \cline{2-6} 
 & Online & $95.22 \pm 4.65^{-1}$ & $95.08 \pm 4.80e^{-1}$ & $95.72 \pm 2.51$ & $94.78 \pm 1.24$ \\ \cline{2-6} 
 & \begin{tabular}[c]{@{}l@{}}Traditional Prot. Learning\end{tabular} & $96.03 \pm 1.43e^{-3}$ & $95.93 \pm 1.44e^{-3}$ & $98.61 \pm 2.54e^{-3}$ & $93.39 \pm 1.12e^{-3}$ \\ \cline{2-6} 
 & \begin{tabular}[c]{@{}l@{}}\textbf{Dual-space Prot. Learning}\end{tabular} & $\textbf{96.22} \pm 6.82e^{-4}$ & $\textbf{96.10} \pm 7.32e^{-4}$ & $\textbf{99.24} \pm 4.80e^{-4}$ & $\textbf{93.16} \pm 1.44e^{-3}$ \\ \cline{1-6} 
\multirow{4}{*}{\textbf{\begin{tabular}[c]{@{}c@{}}MLP W/ \\ Attention\end{tabular}}} & Offline & $99.61 \pm 7.15e^{-2}$ & $99.61 \pm 7.83e^{-2}$ & $99.28 \pm 2.23e^{-1}$ & $99.55 \pm 2.19e^{-1}$ \\ \cline{2-6} 
 & Online & $95.85 \pm 5.37e^{-1}$ & $95.78 \pm 5.98e^{-1}$ & $98.44 \pm 2.31$ & $93.50 \pm 1.82$\\ \cline{2-6} 
 & \begin{tabular}[c]{@{}l@{}}Traditional Prot. Learning\end{tabular} & $96.54 \pm 3.69e^{-3}$ & $96.45 \pm 3.94e^{-3}$ & $97.40 \pm 7.67e^{-4}$ & $93.85 \pm 7.97e^{-3}$\\ \cline{2-6} 
  & \begin{tabular}[c]{@{}l@{}}\textbf{Dual-space Prot. Learning}\end{tabular} & $\textbf{96.44} \pm 1.63e^{-3}$ & $\textbf{96.32} \pm 1.74e^{-3}$ & $\textbf{99.55} \pm 3.72e^{-4}$ & $\textbf{93.30} \pm 3.23e^{-3}$ 
\end{tabular}
\caption{Model Performance - metrics in rates (\%) - whole dataset.}
\label{tab:model-performance}
\end{table*}

\begin{table*}[h]
\centering
\begin{tabular}{c|l|c|c|c|c}
\textbf{Model} & \multicolumn{1}{c|}{\textbf{\begin{tabular}[c]{@{}c@{}}Learning Technique\end{tabular}}} & \textbf{Accuracy} & \textbf{F-1} & \textbf{Precision} & \textbf{Recall} \\ \hline
\multirow{4}{*}{\textbf{MLP}} & Offline & $61.93 \pm 6.02e^{-2}$ & $57.69 \pm 1.17e^{-1}$ & $71.51 \pm 1.53e^{-1}$ & $57.69 \pm 2.73e^{-1}$\\ \cline{2-6} 
 & Online & $55.11 \pm 3.79$ & $53.99 \pm 3.06$ & $54.83 \pm 3.25$ & $52.36 \pm 2.84$ \\ \cline{2-6} 
 & \begin{tabular}[c]{@{}l@{}}Traditional Prot. Learning\end{tabular} & $92.37 \pm 1.90e^{-2}$ & $92.01 \pm 2.25e^{-2}$ & $95.94 \pm 2.05e^{-2}$ & $88.58 \pm 5.24e^{-2}$ \\ \cline{2-6} 
  & \begin{tabular}[c]{@{}l@{}}\textbf{Dual-space Prot. Learning}\end{tabular} & $\textbf{93.87} \pm 7.54e^{-2}$ & $\textbf{93.86} \pm 1.01e^{-1}$ & $\textbf{94.80} \pm 3.82e^{-2}$ & $\textbf{93.44} \pm 1.35e^{-1}$ \\ \cline{1-6} 
\multirow{4}{*}{\textbf{\begin{tabular}[c]{@{}c@{}}MLP W/ \\ Attention\end{tabular}}} & Offline & $65.45 \pm 1.18e^{-1}$ & $65.38 \pm 1.18e^{-1}$ & $74.01 \pm 1.92e^{-1}$ & $70.89 \pm 2.31e^{-1}$ \\ \cline{2-6} 
 & Online & $58.46 \pm 2.16$ & $57.51 \pm 2.28$ & $58.78 \pm 3.10$ & $54.94 \pm 1.91$\\ \cline{2-6} 
 & \begin{tabular}[c]{@{}l@{}}Traditional Prot. Learning\end{tabular} & $94.05 \pm 8.72e^{-3}$ & $93.91 \pm 7.91e^{-3}$ & $96.37 \pm 2.79e^{-2}$ & $91.64 \pm 1.83e^{-2}$\\ \cline{2-6} 
 & \begin{tabular}[c]{@{}l@{}}\textbf{Dual-space Prot. Learning}\end{tabular} & $\textbf{94.85} \pm 2.06e^{-2}$ & $\textbf{94.71} \pm 2.43e^{-2}$ & $\textbf{97.30} \pm 1.41e^{-2}$ & $\textbf{92.31} \pm 4.72e^{-2}$ 
\end{tabular}
\caption{Model Performance - metrics in rates (\%) - training $N$: 100.}
\label{tab:model-performance-reduced}
\end{table*}

The first graph showcases the Offline Training with Cross entropy loss, where both Regular MLP and MLP with Attention demonstrate a rapid convergence, with the attention model slightly underperforming in this context. The Online Training graph shows a higher variability in validation loss, particularly for the attention model, suggesting that the online setting may introduce complexities that are not as effectively managed by the attention mechanism within the constraints of a small dataset.

Shifting to the Prototypical Loss graphs, we see that the validation loss decreases steadily for both models, yet the addition of attention does not significantly diverge from the performance of the regular model, implying that the prototypical loss alone is a strong contributor to the learning process.

Most notably, the offline training with our dual-space loss graph reveals that the dual-space prototypical network, especially when coupled with attention, achieves a lower validation loss much quicker than the other methods. This supports our hypothesis that the dual-space approach, which leverages both the magnitude of feature differences and their directional alignment in the latent space, provides a more robust learning mechanism. Including attention seems to amplify this effect, allowing for a more focused and discriminative feature learning, which is critical when working with sparse datasets.

These results affirm the strength \textit{D-space} in offline settings, where it can leverage the full extent of the dataset. It's an encouraging indication that our model is well-suited for scenarios with limited data availability, making it a promising candidate for further exploration in few-shot learning and other data-constrained environments.

\subsection{Model Performance}

This section analyzes our models' fare in detecting DDoS threats, comparing traditional MLPs against attention-enhanced variants under different learning conditions. When trained on the full dataset and the limited set from Section \ref{subsec:training-process-analysis}, we evaluate their accuracy and reliability. To ensure statistical significance, we ran 30 experiments per configuration (Section \ref{expsetup}).

The results from Table \ref{tab:model-performance} highlight the robust performance of the traditional MLP models, particularly when augmented with an attention mechanism. The offline training paradigm, which encompasses the entire dataset, allows these models to extract comprehensive patterns, a capability reflected in the exceptional accuracy and F-1 scores that exceed 99\% for the attention-enhanced MLP. Notably, these models optimally perform when provided with the full spectrum of data, enabling them to establish complex relationships and predictive capabilities crucial for effective DDoS threat detection.

While not surpassing the regular MLP in a full-data offline training context, the prototypical learning approaches show their true potential when data is scarce. They are designed to generalize from minimal examples and are adaptable, making them particularly suitable for real-world applications with prevalent data limitations. This design intention is validated in Table \ref{tab:model-performance-reduced}, where we observe a decline in performance metrics for both the standard MLP and the MLP with Attention under limited data training. The accuracy and F-1 scores dip to around 60\%, revealing the challenges posed by data scarcity.

\textit{D-space} outperforms the baselines with an average accuracy of 94.85\% and consistent F-1 scores, exhibiting minimal variance across runs. This resilience demonstrates the efficacy of the dual-space approach, which synthesizes learning from both feature and label spaces to form a robust representation of each class, even from limited data samples. The precision metric, where the dual-space prototypical network achieves an average of 99.55\% precision across all tests, underscores the model's ability to pinpoint true positive threats with high reliability.


\section{Conclusion} \label{sec:conclusion}

This paper introduced the Dual-space loss function and the Dual-space Prototypical Network, advancements engineered for detecting Distributed Denial of Service attacks. Grounded in the principles of prototypical networks and enhanced by our novel loss function, these innovations mark a significant leap forward in cybersecurity. The dual-space Loss function, in particular, is designed to leverage geometric and angular metrics, enabling a more refined and adaptable approach to understanding the latent space of data representations.

Our approach stands out when utilized on neural networks trained on small datasets, a critical capability given the dynamic and often data-scarce landscape of cybersecurity threats. In scenarios where traditional models might falter due to insufficient training samples, \textit{D-space} thrives, maintaining high accuracy and F1 scores. This performance shows the model's deep understanding of the underlying data structure, allowing it to extract meaningful insights from minimal input. 

In the future, our focus will pivot towards further enhancing the adaptability and efficiency of our dual-space prototypical network, pushing the envelope of representation learning within the cybersecurity domain. By exploring advanced techniques in representation learning, we envision creating models that can learn more effectively and uncover more profound insights into attacks. We aim to remain at the forefront of the battle against cybersecurity threats, designing proactive solutions and ensuring a safer digital environment for all.


\bibliographystyle{IEEEtran}
\bibliography{refs}

\begin{thebibliography}{10}
\providecommand{\url}[1]{#1}
\csname url@samestyle\endcsname
\providecommand{\newblock}{\relax}
\providecommand{\bibinfo}[2]{#2}
\providecommand{\BIBentrySTDinterwordspacing}{\spaceskip=0pt\relax}
\providecommand{\BIBentryALTinterwordstretchfactor}{4}
\providecommand{\BIBentryALTinterwordspacing}{\spaceskip=\fontdimen2\font plus
\BIBentryALTinterwordstretchfactor\fontdimen3\font minus \fontdimen4\font\relax}
\providecommand{\BIBforeignlanguage}[2]{{%
\expandafter\ifx\csname l@#1\endcsname\relax
\typeout{** WARNING: IEEEtran.bst: No hyphenation pattern has been}%
\typeout{** loaded for the language `#1'. Using the pattern for}%
\typeout{** the default language instead.}%
\else
\language=\csname l@#1\endcsname
\fi
#2}}
\providecommand{\BIBdecl}{\relax}
\BIBdecl

\bibitem{guo2023review}
Y.~Guo, ``A review of machine learning-based zero-day attack detection: Challenges and future directions,'' \emph{Computer Communications}, vol. 198, pp. 175--185, 2023.

\bibitem{wei2022abl}
W.~Wei, H.~Gu, W.~Deng, Z.~Xiao, and X.~Ren, ``Abl-tc: A lightweight design for network traffic classification empowered by deep learning,'' \emph{Neurocomputing}, vol. 489, pp. 333--344, 2022.

\bibitem{debicha2023tad}
I.~Debicha, R.~Bauwens, T.~Debatty, J.-M. Dricot, T.~Kenaza, and W.~Mees, ``Tad: Transfer learning-based multi-adversarial detection of evasion attacks against network intrusion detection systems,'' \emph{Future Generation Computer Systems}, vol. 138, pp. 185--197, 2023.

\bibitem{mirkovic2004taxonomy}
J.~Mirkovic and P.~Reiher, ``A taxonomy of ddos attack and ddos defense mechanisms,'' \emph{ACM SIGCOMM Computer Communication Review}, vol.~34, no.~2, pp. 39--53, 2004.

\bibitem{kolias2017ddos}
C.~Kolias, G.~Kambourakis, A.~Stavrou, and J.~Voas, ``Ddos in the iot: Mirai and other botnets,'' \emph{Computer}, vol.~50, no.~7, pp. 80--84, 2017.

\bibitem{newman20181}
L.~H. Newman, ``a 1.3-tbs ddos hit github, the largest yet recorded,'' \emph{Recuperado de https://www. wired. com/story/github-ddos-memcached}, 2018.

\bibitem{rahouti2022sdn}
M.~Rahouti, K.~Xiong, Y.~Xin, S.~K. Jagatheesaperumal, M.~Ayyash, and M.~Shaheed, ``\mbox{SDN} security review: Threat taxonomy, implications, and open challenges,'' \emph{IEEE Access}, vol.~10, pp. 45\,820--45\,854, 2022.

\bibitem{dhillon2021building}
H.~Dhillon, ``Building effective network security frameworks using deep transfer learning techniques,'' Ph.D. dissertation, The University of Western Ontario (Canada), 2021.

\bibitem{David2021Discriminating}
J.~David and C.~Thomas, ``Discriminating flash crowds from \mbox{DDoS} attacks using efficient thresholding algorithm,'' \emph{JPDC}, vol. 152, pp. 79--87, 2021, Elsevier.

\bibitem{Chin2018Kernel}
T.~Chin, K.~Xiong, and M.~Rahouti, ``Kernel-space intrusion detection using software-defined networking,'' \emph{EAI Endorsed Transactions on Security and Safety}, vol.~5, no.~15, p.~e2, 2018.

\bibitem{Zheng2018Realtime}
J.~Zheng, Q.~Li, G.~Gu, J.~Cao, m.~Yau, and J.~Wu, ``Realtime \mbox{DDoS} defense using \mbox{COTS SDN} switches via adaptive correlation analysis,'' \emph{TIFS}, vol.~13, no.~7, pp. 1838--1853, 2018, IEEE.

\bibitem{Rahouti2021SynGuard}
M.~Rahouti, K.~Xiong, N.~Ghani, and F.~Shaikh, ``\mbox{SYNGuard}: Dynamic threshold-based \mbox{SYN} flood attack detection and mitigation in software-defined networks,'' \emph{IET Networks}, vol.~10, no.~2, pp. 76--87, 2021.

\bibitem{Liang2019Empirical}
X.~Liang and T.~Znati, ``An empirical study of intelligent approaches to \mbox{DDoS} detection in large scale networks,'' in \emph{ICNC}.\hskip 1em plus 0.5em minus 0.4em\relax IEEE, 2019, pp. 821--827.

\bibitem{owusu2023enhancing}
E.~Owusu, M.~Rahouti, D.~F. Hsu, K.~Xiong, and Y.~Xin, ``Enhancing ml-based dos attack detection through combinatorial fusion analysis,'' in \emph{2023 IEEE Conference on Communications and Network Security (CNS)}.\hskip 1em plus 0.5em minus 0.4em\relax IEEE, 2023, pp. 1--6.

\bibitem{alfatemi2024advancing}
A.~Alfatemi, M.~Rahouti, R.~Amin, S.~ALJamal, K.~Xiong, and Y.~Xin, ``Advancing ddos attack detection: A synergistic approach using deep residual neural networks and synthetic oversampling,'' \emph{arXiv preprint arXiv:2401.03116}, 2024.

\bibitem{zhang2019network}
Y.~Zhang, X.~Chen, L.~Jin, X.~Wang, and D.~Guo, ``Network intrusion detection: Based on deep hierarchical network and original flow data,'' \emph{IEEE Access}, vol.~7, pp. 37\,004--37\,016, 2019.

\bibitem{zhong2020helad}
Y.~Zhong, W.~Chen, Z.~Wang, Y.~Chen, K.~Wang, Y.~Li, X.~Yin, X.~Shi, J.~Yang, and K.~Li, ``Helad: A novel network anomaly detection model based on heterogeneous ensemble learning,'' \emph{Computer Networks}, vol. 169, p. 107049, 2020.

\bibitem{li2022mfvt}
M.~Li, D.~Han, D.~Li, H.~Liu, and C.-C. Chang, ``Mfvt: an anomaly traffic detection method merging feature fusion network and vision transformer architecture,'' \emph{EURASIP Journal on Wireless Communications and Networking}, vol. 2022, no.~1, p.~39, 2022.

\bibitem{sun2023few}
H.~Sun, L.~Wan, M.~Liu, and B.~Wang, ``Few-shot network intrusion detection based on prototypical capsule network with attention mechanism,'' \emph{Plos one}, vol.~18, no.~4, p. e0284632, 2023.

\bibitem{miao2023spn}
G.~Miao, G.~Wu, Z.~Zhang, Y.~Tong, and B.~Lu, ``Spn: A method of few-shot traffic classification with out-of-distribution detection based on siamese prototypical network,'' \emph{IEEE Access}, 2023.

\bibitem{hosseini2023intrusion}
S.~Hosseini, ``Intrusion detection in \mbox{IoT} network using few-shot class incremental learning,'' Ph.D. dissertation, Carleton University, 2023.

\bibitem{tian2022few}
J.-Y. Tian, Z.-M. Wang, H.~Fang, L.-M. Chen, J.~Qin, J.~Chen, Z.-H. Wang \emph{et~al.}, ``Few-shot learning-based network intrusion detection through an enhanced parallelized triplet network,'' \emph{Security and Communication Networks}, vol. 2022, 2022.

\bibitem{iliyasu2022few}
A.~S. Iliyasu, U.~A. Abdurrahman, and L.~Zheng, ``Few-shot network intrusion detection using discriminative representation learning with supervised autoencoder,'' \emph{Applied Sciences}, vol.~12, no.~5, p. 2351, 2022.

\bibitem{yang2022fs}
J.~Yang, H.~Li, S.~Shao, F.~Zou, and Y.~Wu, ``Fs-ids: A framework for intrusion detection based on few-shot learning,'' \emph{Computers \& Security}, vol. 122, p. 102899, 2022.

\bibitem{yu2020intrusion}
Y.~Yu and N.~Bian, ``An intrusion detection method using few-shot learning,'' \emph{IEEE Access}, vol.~8, pp. 49\,730--49\,740, 2020.

\bibitem{kwon2018empirical}
D.~Kwon, K.~Natarajan, S.~C. Suh, H.~Kim, and J.~Kim, ``An empirical study on network anomaly detection using convolutional neural networks,'' in \emph{2018 IEEE 38th International Conference on Distributed Computing Systems (ICDCS)}.\hskip 1em plus 0.5em minus 0.4em\relax IEEE, 2018, pp. 1595--1598.

\bibitem{wang2018network}
W.~Wang, Y.~Bai, C.~Yu, Y.~Gu, P.~Feng, X.~Wang, and R.~Wang, ``A network traffic flow prediction with deep learning approach for large-scale metropolitan area network,'' in \emph{NOMS 2018-2018 IEEE/IFIP Network Operations and Management Symposium}.\hskip 1em plus 0.5em minus 0.4em\relax IEEE, 2018, pp. 1--9.

\bibitem{jiang2020network}
K.~Jiang, W.~Wang, A.~Wang, and H.~Wu, ``Network intrusion detection combined hybrid sampling with deep hierarchical network,'' \emph{IEEE access}, vol.~8, pp. 32\,464--32\,476, 2020.

\bibitem{liu2021fast}
J.~Liu, Y.~Gao, and F.~Hu, ``A fast network intrusion detection system using adaptive synthetic oversampling and lightgbm,'' \emph{Computers \& Security}, vol. 106, p. 102289, 2021.

\bibitem{li2019method}
J.~Li, X.~Yun, M.~Tian, J.~Xie, S.~Li, Y.~Zhang, and Y.~Zhou, ``A method of http malicious traffic detection on mobile networks,'' in \emph{2019 IEEE Wireless Communications and Networking Conference (WCNC)}.\hskip 1em plus 0.5em minus 0.4em\relax IEEE, 2019, pp. 1--8.

\bibitem{xu2020method}
C.~Xu, J.~Shen, and X.~Du, ``A method of few-shot network intrusion detection based on meta-learning framework,'' \emph{IEEE Transactions on Information Forensics and Security}, vol.~15, pp. 3540--3552, 2020.

\bibitem{snell2017prototypical}
J.~Snell, K.~Swersky, and R.~S. Zemel, ``Prototypical networks for few-shot learning,'' 2017.

\bibitem{pittir}
S.~T. Zargar, J.~Joshi, and D.~Tipper, ``A survey of defense mechanisms against distributed denial of service (ddos) flooding attacks,'' \emph{IEEE Communications Surveys \& Tutorials}, vol.~15, no.~4, pp. 2046--2069, 2013.

\bibitem{pepernot}
N.~Papernot, P.~McDaniel, A.~Sinha, and M.~Wellman, ``Towards the science of security and privacy in machine learning,'' 2016.

\bibitem{sharafaldin2018toward}
I.~Sharafaldin, A.~H. Lashkari, and A.~A. Ghorbani, ``Toward generating a new intrusion detection dataset and intrusion traffic characterization,'' in \emph{ICISSP}, Portugal, January 2018.

\bibitem{CICFlowMeter}
C.~I. for Cybersecurity~(CIC), ``Cicflowmeter: Network traffic flow generator tool,'' \url{https://www.unb.ca/cic/datasets/ids-2017.html}, 2017.

\bibitem{kim2017structured}
Y.~Kim, C.~Denton, L.~Hoang, and A.~M. Rush, ``Structured attention networks,'' 2017.

\bibitem{kingma2017adam}
D.~P. Kingma and J.~Ba, ``Adam: A method for stochastic optimization,'' 2017.

\end{thebibliography}

\end{document}